\documentclass[11pt]{article}
\usepackage{moriond,epsfig}
\usepackage{times}
\usepackage{amsmath}

\input epsf
\epsfverbosetrue

\newcommand{\gtequiv}{\lower2pt\hbox{$\:\stackrel{>}{
            \scriptstyle\sim}\:$}}
\newcommand{\ltequiv}{\lower2pt\hbox{$\:\stackrel{<}{
            \scriptstyle\sim}\:$}}

\begin{document}

\title{DOPING STATE OF MULTI-WALL CARBON NANOTUBE WIRES AND QUANTUM DOTS}

\author{\underline{Christian Sch{\"o}nenberger}, Mark Buitelaar,
Michael Kr{\"u}ger, Isabelle Widmer, Thomas Nussbaumer, and Mahdi Iqbal}

\address{Institute of Physics, University of Basel, Klingelbergstrasse 82,
CH-4056 Basel, Switzerland}

\maketitle

\begin{abstract}
Employing strong electrostatic gating (liquid-ion gating),
the position of the Fermi energy $E_F$ (relative to the charge-neutrality
point) was determined
in multi-wall carbon nanotubes (MWNTs).
$E_F$ is negative (hole doping) and amounts to
\mbox{$\approx -0.3$\,eV} for MWNTs in air.
Evidence that water, and not oxygen, is the main source
of doping has been found.
As a consequence, the number $M$ of occupied $1$d-modes
(not counting spin) is $>2$, i.e. \mbox{$M\approx 10$}.
This is supported by the single-electron
level spacing, deduced from observed single-electron charging
effects (SET) at low temperature.
The latter are dominated by co-tunneling processes, as we
observe the Kondo effect. This provides evidence, that
highly transmissive channels are present in MWNTs
at low temperature, despite the `disorder' observed
in previous experiments.
\end{abstract}

\section{Introduction}

Carbon nanotubes (CNTs) are ideal model
systems for the exploration of
electrical transport in low dimensions.\cite{PhysWorld}
Two sorts of nanotubes (NTs) exist:
Single-wall and multi-wall NTs. An ideal (undoped)
SWNT can be either metallic or semiconducting, but
here, we only focus on metallic NTs. In these
tubes, the current is carried by two
modes,\footnote{\,If we refer to $M$ modes, spin-degeneracy
is not included.} leading
to a conductance of $4e^2/h$, provided
backscattering is absent.
Recent experiments have shown that
scattering within metallic SWNT is weak.
In a particular nice experiment the electrostatic potential drop
along a voltage-biased NT was measured by using the tip of an
atomic-force microscope as a probe.\cite{BachtoldAFM}
For SWNTs, most of the
potential drops at the contacts.
In contrast, for MWNTs a considerable fraction of the potential drops
along the tube, suggesting intrinsic scattering in MWNTs.
A length dependent resistance was deduced before from
electric resistance measurements on multiply contacted MWNTs.\cite{ApplPhysA}
The typical value for the resistance per unit length
is \mbox{$R^{\prime}=5-10$\,k$\Omega/\mu m$}.
We mention, that there is one conflicting results:
Frank {\it et al.}\cite{Frank98} came to the conclusion
that MWNTs are ballistic conductors
even at room temperature.\footnote{\,Note,
that ballistic transport
is not expected in MWNTs at room temperature, because the
energy-separation between $1$d subbands is comparable to $kT$.}
Seemingly compelling evidence for diffusive transport in MWNTs
is provided by measurements of the magnetoresistance, both
in parallel and perpendicular magnetic field.\cite{ApplPhysA}
For example, the resistance
modulation in parallel magnetic field can be described very well
by the Altshuler-Aronov-Spivak (AAS) theory (weak-localization
in a cylindrical conductor), which relies
on {\em diffusive} transport.\cite{BachtoldAB}
These experiments did also show that
the electrical current is preferentially carried by the
outermost tube, at least at low temperatures.
Hence, a single nanotube is probed, albeit one
with a large diameter of \mbox{$d\approx 10-20$\,nm},
which is about ten times larger than that of prototype SWNTs.
As emphasized before, a metallic SWNT is characterized
by only $M=2$ $1$d-modes, a property that should
be independent of the diameter.
How can we than reconcile
the availability of only $2$ modes for an ideal NT
with the observation of diffusive motion.
Diffusive transport requires $M \gg 1$. May it be that
MWNTs are doped to a such a degree that $M \gg 1$?

By using a new gating technique (electrochemical gating),
we have recently shown that MWNTs are indeed (hole-) doped.\cite{KruegerAPL}
The number of $1$d-modes is $\not = 2$, but rather
$10-20$, see section~2. MWNTs are not single-mode,
but rather few mode quasi-one-dimensional
wires. Whether they are $1$d diffusive, i.e. quasi-ballistic
with a mean-free path $l$ exceeding the circumference $\pi d$,
or $2d$ diffusive \mbox{($l \leq \pi d$)} is another question.

Taking \mbox{$R^{\prime}=(2Me^2l/h)^{-1}=5$\,k$\Omega/\mu$}
($L$ is tube length),
yields \mbox{$l\approx 100$\,nm} which is
of order of the circumference. This simple estimate
is in good agreement with measurements of the
energy-dependent tunneling DOS $\nu(E)$.
$\nu(E)$ is not structureless,
as would be expected if \mbox{$l \ll \pi d$}, but shows
features reminiscent of quantization into $1$d-modes,
albeit with some broadening.\cite{Bachtold_condmat}


Recently, we have studied gate-induced conductance fluctuations
in MWNTs at low temperatures and tried to compare the measurements
with UCF theory. In the regime of thermally-induced averaging,
i.e. for tubes which are much longer than
the phase-coherence length $l_{\phi}$ and/or
the thermal length $l_T$,
the functional dependence is in agreement with theory.
These data allow to deduce $l_{\phi}$, which
follows Nyquist-dephasing below \mbox{$\approx 4$\,K}.
However, when we approach the universal limit,
i.e. if  $L \approx l_{\phi}$, the
temperature dependence of conductance fluctuations
markedly deviates from standard theories.
This has led us to study
shorter tubes in more detail by measuring
the differential conductance $G_d=dI/dV$
as a function of transport voltage $V$
and gate voltage $U_g$ in the
fully coherent regime, i.e. for $l_{\phi}\sim L$.
Displaying $G_d(U_g,V)$ in a greyscale plot
helps to recognize the underlying physics. This is
in particular true for single-electron charging
effects which might be present simultaneously to
quantum interference effects, both modulating the
equilibrium conductance.

Single-electron charging effects
(single-electron tunneling = SET), such as
Coulomb blockade and Coulomb oscillations were
observed in SWNTs from the beginning.\cite{CB_SWNT}
However, in our own work on MWNTs we have
never observed clear evidence of Coulomb blockade until now.
We have argued that this absence is due to the low-ohmic
contacts in our experiments, which are always
of order \mbox{$\ltequiv 10$\,k$\Omega$}.\cite{ApplPhysA}
In contrast, measurements on MWNTs with high-ohmic
contacts (\mbox{$> 10^6$\,$\Omega$}) display
the conventional features of single-electron
charging effects.\cite{Hakonen}
The evaporation of Au over the nanotubes, the method
we prefer for fabricating contacts, leads to
contact resistances that can be as low as
\mbox{$1$\,k$\Omega$} at room temperature.
This is low enough to suppress SET. At cryogenic temperature, however,
contact resistances usually increase, so that
SET may show up. In section~3 we present our first (and still rare)
observation of SET in transport through a MWNT
with `low-ohmic' contacts.
Since the coupling to the contacts
is rather strong, the conductance is dominated by higher-order
co-tunneling processes. These new data
allow to extract the single-particle level spacing $\delta E$
of the MWNT quantum dot. Similar to the result from electrochemical
gating, the measured $\delta E$ suggests that $\approx 10$ modes
are occupied.

\section{Doping state of multi-wall carbon nanotubes}

\begin{figure}[h]
\epsfxsize=90mm \centerline{\epsfbox{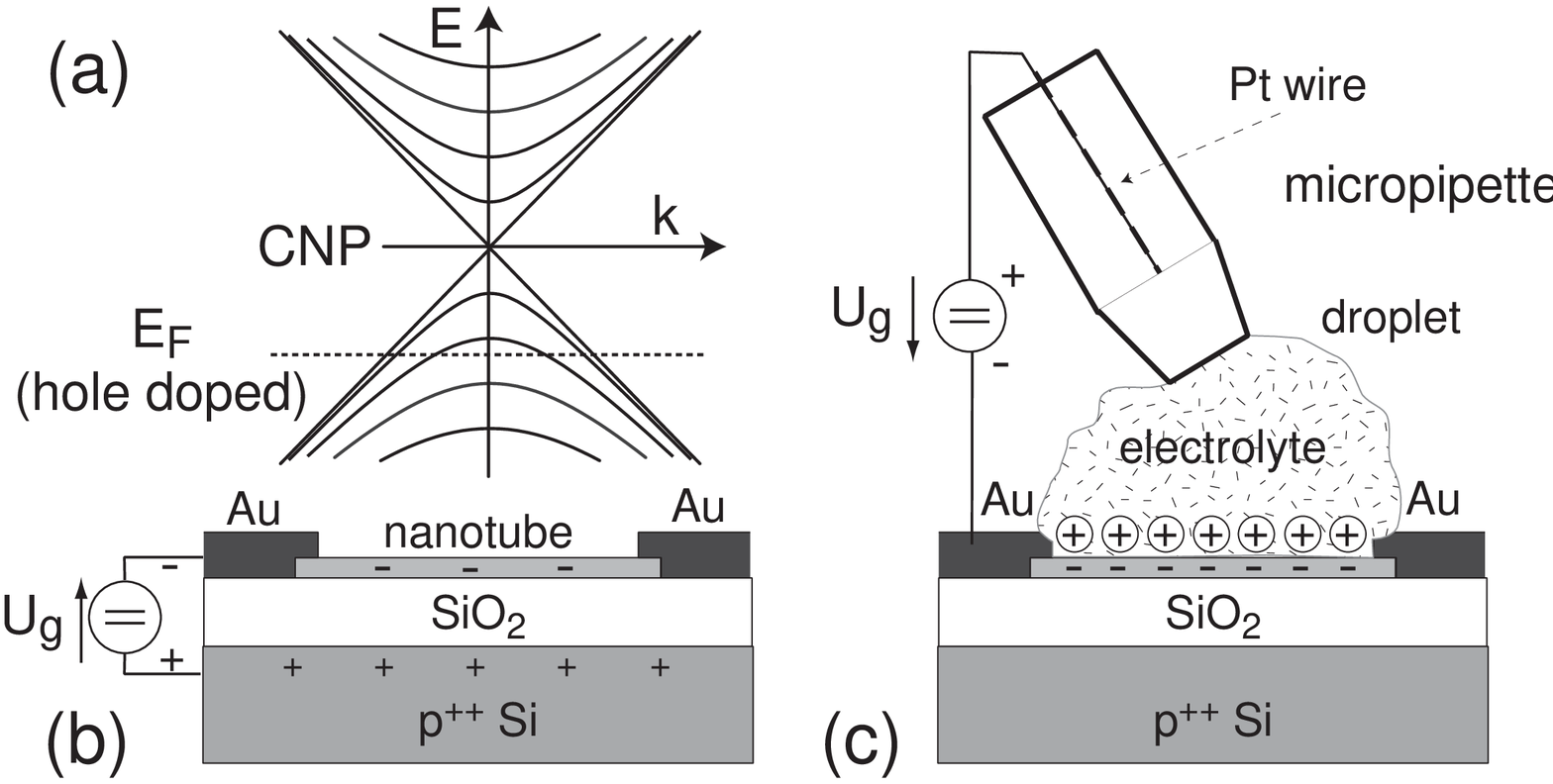}} \caption{
  (a) Simplified $1$d bandstructure of a metallic carbon nanotube.
  CNP, the charge neutrality point, refers to the position of the
  Fermi energy $E_F$ of an undoped nanotube. The nanotube is hole (electron) doped
  if $E_F < CNP$ ($E_F > CNP$). $E_F$ can be changed via the gate voltage
  $U_g$, conventionally applied to the substrate (backgate), as shown in (b).
  Much larger Fermi-energy shifts are possible through liquid-ion gating, (c).
  \mbox{LiClO$_4$} ($1-500$\,mM) was used as electrolyte.
  }
\end{figure}

In order to determine the degree of doping in MWNTs
(or other nanotubes) the position of the Fermi
energy need to be determined. One possible approach
is to measure the conductance as a function of a gate
voltage $U_g$, which shifts the Fermi energy $E_F$.
If the conductance $G$ increases
(decreases) with increasing $E_F$, the NT is
n-doped (p-doped), see Fig.~1a.
At, or in the vicinity of the charge-neutrality
point (CNP), the number of modes is minimal.
It is $2$ for a metallic tube and zero for a semiconducting
one.\footnote{\,At room temperature the $1$d bandstructure
in Fig.~1a is considerably smeared, so that no abrupt
transitions are expected if $E_F$ crosses a band onset.}
Hence, one expects to see a conductance minimum (or
a resistance maximum) if $E_F$ crosses the CNP.
Ideally, for an undoped NT, $E_F$ should lie
at the CNP and the resistance maximum should
appear at $U_g=0$. If this maximum is observed
at $U_g \not = 0$, the Fermi level is shifted due
to some doping. In order to deduce $E_F$ from
the measured gate voltage, the electrochemical
capacitance between gate and NT needs to be known.
Conventional back gates (see Fig.~1b)
have relatively low capacitance. Consequently
Fermi-level shifts are small even for large applied
gate voltages. In this case, the resistance maximum
cannot be measured, and hence, the doping level cannot
be determined. This problem is circumvented in our
approach, in which the coupling capacitance is
made very large. We immerse the
nanotube into an electrolyte through which
gating is achieved, see Fig~1c. For details, see
Ref.~6.
Very large gate capacitances are possible due to
the large double-layer capacitance of the electrolyte
NT interface. This has an
important implication: Because the double-layer
capacitance is much larger than the electrochemical
capacitance of the NT, the total capacitance
(series connection of the two) is
determined by the nanotube alone. Hence, there
is a one-to-one correspondence between gate-voltage
change $\Delta U_g$ and Fermi level shift $\Delta E_F$,
i.e. $\Delta E_F\cong\Delta U_g$.

Electrochemical gating is studied at room temperature
on single MWNTs with lithographically defined Au contacts
evaporated over the NTs.
The nanotube-contact structure is fabricated on
degenerately doped Si with a \mbox{$400$\,nm}
thick \mbox{SiO$_2$} spacer layer. The subtsrate
can be used as a conventional backgate, Fig~1b.
If, as outlined in Fig.~1c,
a positive gate voltage $U_g$ is applied,
the NT-electrolyte interface is polarized
by the attraction of cations. Consequently,
the Fermi level increases in the NT to
maintain charge neutrality.

An example for the dependence of the nanotube
resistance on (electrochemical) gate voltage is shown in Fig.~2.
A pronounced resistance maximum is seen
at $U_g = 1V$, which we identify as the CNP.
In order to reach the CNP the Fermi energy must
first be raised by \mbox{$\approx 1$\,eV}. This
suggests that the NT is considerably hole-doped at $U_g=0$.
There is, however, a time dependence (not shown).\cite{KruegerAPL}
After immersion, the resistance maximum is observed
at a smaller voltage of \mbox{$\approx 0.3$\,eV}. The
position of maximum $R$ then gradually moves to larger
voltages to finally stabilize at \mbox{$1$\,V} in
this example. Hence, the doping level increases in the electrolyte.
Typical Fermi-level shifts in air (before immersion)
amount to \mbox{$0.3-0.5$\,eV}.

\begin{figure}
\epsfxsize=160mm
\centerline{\epsfbox{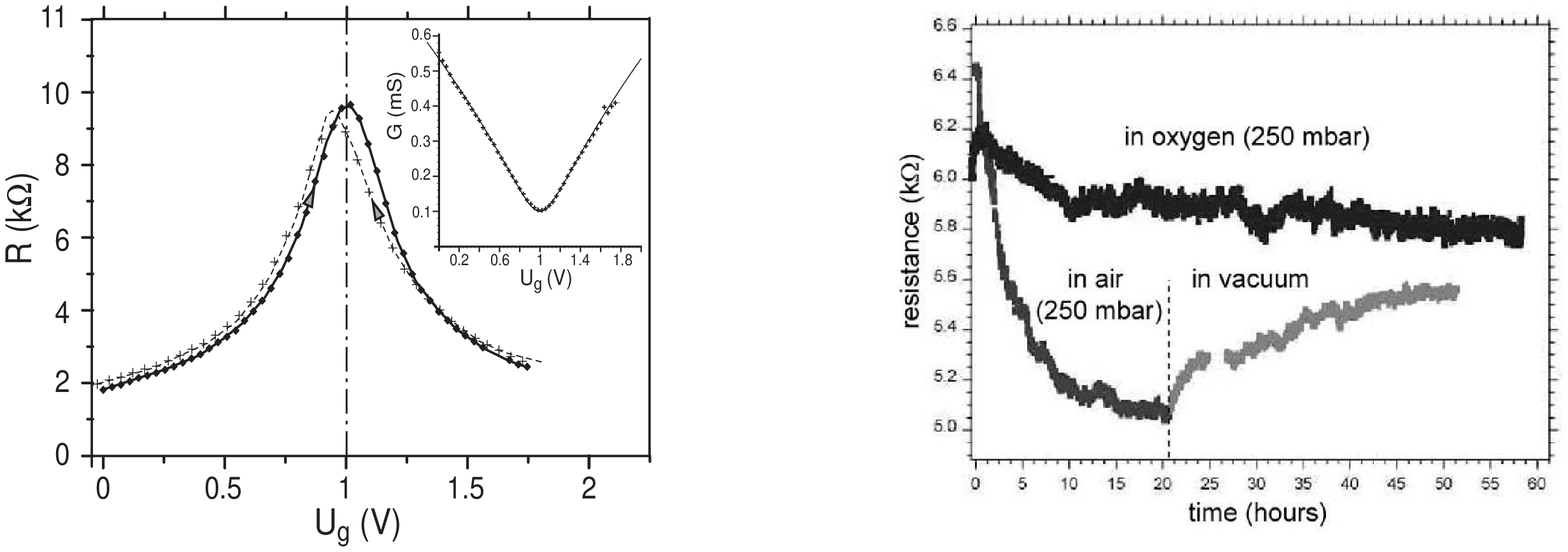}}
\caption{(Left) The typical evolution of the electrical resistance $R$
  of a single MWNT with gate voltage $U_g$
  measured in a \mbox{$10$\,mM} \mbox{LiClO$_4$}
  electrolyte.
  Inset: Comparison of $G=1/R$ ($\bullet$)
  with theory (full curve), see Ref.~6.
  }
\caption{(Right) Doping induced change of the electrical resistance
  caused by the adsorption of (a) oxygen and (b) air. The nanotube
  was annealed before exposure. The resistance
  decrease is caused by increased hole-doping. Air
  (most likely water vapour) has a much bigger effect
  than pure oxygen.
  }
\end{figure}

Based on measurements of the electrical resistance of SWNTs
in a controlled environment of different gases it was
suggested that hole-doping is induced by
by oxygen.\cite{gasexp} Our own experiments show that oxygen has an effect
on the doping state of MWNTs. However, the effect is
weak, as much larger doping shifts are seen in
ambient air. An example is shown in Fig.~3.
Here, we compare the evolution of the resistance
of a MWNT at zero (back-) gate
voltage when exposed to oxygen and air. Prior to
exposure, the NT was
annealed at \mbox{$50$\,C$^{\circ}$} in vacuum.
As a consequence of added hole doping the resistance
is seen to decrease. The decrease is much more pronounced
for air than for oxygen. The resistance
increases again in vacuum, nicely demonstrating that
the oxidizing species are only weakly adsorbed.
The strongest source of doping is most likely due to water,
as large Fermi-level shifts can be induced by water vapour
alone.

For the interpretation of
previous electrical measurements,
the net doping concentration $Q_d$
and the Fermi-level shift for
a `virgin' MWNT in air
are important.
Typical values for the latter
are \mbox{$0.3-0.5$\,eV}.
Comparing this with the average $1$d subband spacing
\mbox{$\hbar v_F/2d$} (\mbox{$\approx 33$\,meV}
for a \mbox{$10$\,nm} diameter NT),
we conclude that
$M=9-15$ subbands may contribute to the conductance
instead of $2$ for an ideal metallic NT.
A doping-induced \mbox{$E_F=0.3$\,eV}
corresponds to a doping concentration of
\mbox{$Q_d^{\prime}/e\approx 2\cdot 10^3$\,$\mu$m$^{-1}$}, or
expressed per surface area to
\mbox{$Q_d/e\approx 0.7\cdot 10^{13}$\,cm$^{-2}$}
giving approximately one elementary charge
per $500$ carbon atoms.

\section{Multi-wall carbon nanotubes as quantum dots}

As mentioned in the introduction single-electron tunneling (SET)
such as Coulomb oscillations have been observed in MWNTs
and SWNTs, provided the contacts were high-ohmic, i.e.
\mbox{$> 1$\,M$\Omega$}.\cite{CB_SWNT,Hakonen}
In the opposite limit of relatively low-ohmic contacts
of order \mbox{$\ltequiv 10$\,k$\Omega$} charging effects have
not yet been observed in transport through MWNTs.
Though we have observed pronounced conductance oscillations
as a function of gate voltage~\cite{Buitelaar}
and also as a function of
magnetic field in the latter type of samples~\cite{ApplPhysA}, these cannot readily be assigned
to charging effects, as the oscillations were strongly aperiodic.
We have assigned these oscillations
to quantum interference of non-interacting quasi-particles,
with the interference being caused by the presence of randomly
distributed scatterers. It may be that this
interpretation need to be reconsidered, since we have recently
observed SET effects in MWNTs low-ohmic contacts (\mbox{$\ltequiv 10$\,k$\Omega$}).

In order to entangle SET from quantum interference it is
useful to measure $dI/dV$ as a function of gate voltage $U_g$
and transport voltage $V$. The obtained greyscale plots
greatly helps to recognize the underlying physics.

\begin{figure}[h]
  \epsfxsize=140mm
  \centerline{\epsfbox{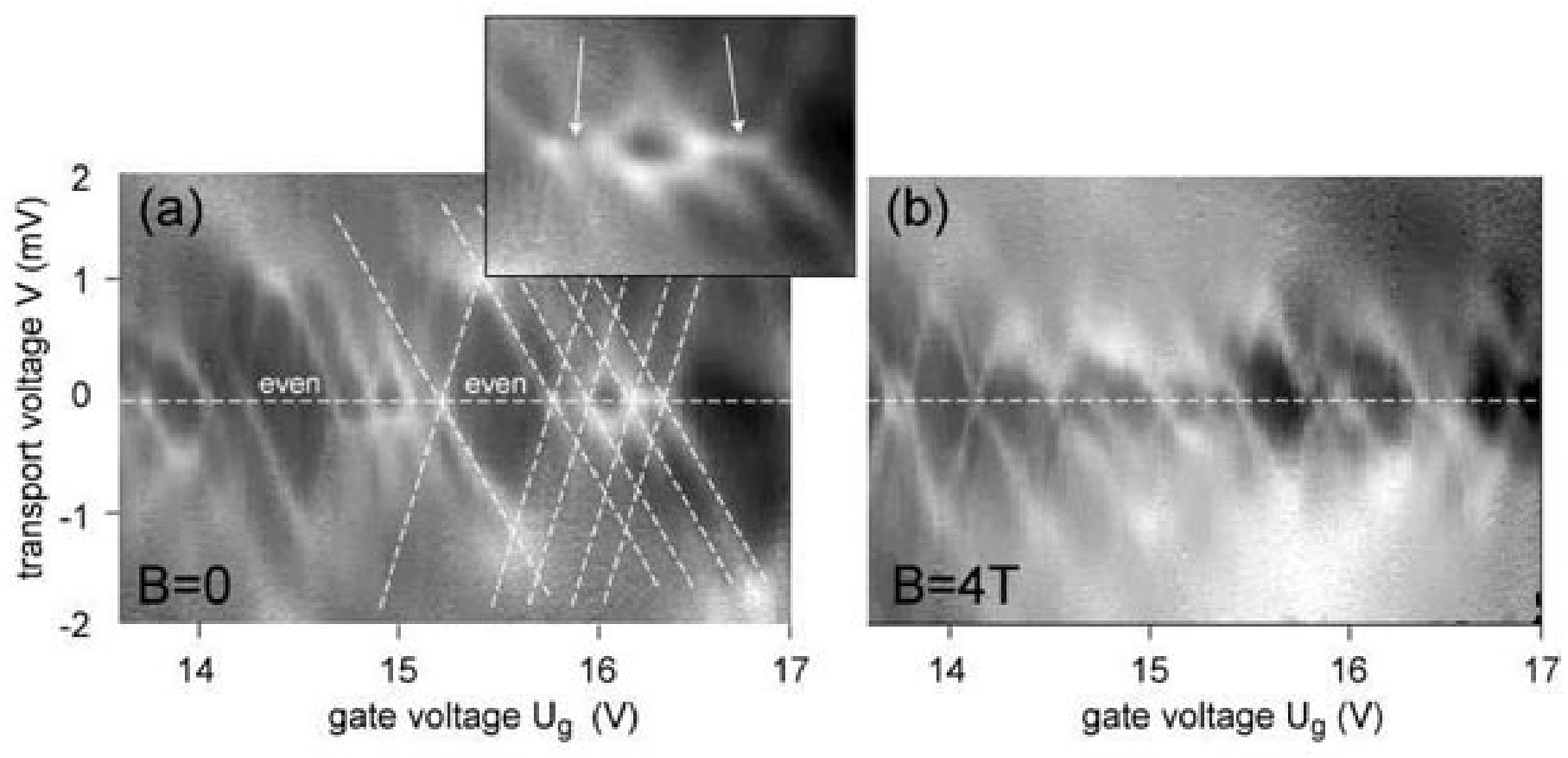}}
  \caption{Differential conductance $dI/dV$
  as a function of transport voltage $V$ (vertical) and
  gate voltage $V_g$ (horizontal). White (black) corresponds to high (low)
  conductance. (a) was measured in zero magnetic field $B$ and
  (b) in \mbox{$B=4$\,T}. The observed pattern is caused by
  single-electron charging effects. The inset shows a zoomed in portion
  of (a) in which two horizontal ridges (arrows) of high conductance are
  seen. These ridges are caused by the Kondo effect.
}
\end{figure}

In small (semi) isolated objects,
like our MWNT devices, transport may be blocked by
Coulomb interaction (Coulomb blockade). This is the case,
if the energy for adding a single electron to the tube
(the single-electron charging energy $e^2/2C$) is larger
than the thermal energy $kT$ and if the object is only
weakly coupled to external leads (contact resistance
$\gtequiv h/2e^2$). Coulomb blockade is suppressed
if the energy of two charge states are degenerate. For example,
if $E(N)=E(N+1)$, where $E(N)$ denotes the total ground-state
energy of the NT with $N$ electrons, an equilibrium current
can flow via the sequence $N\rightarrow N+1 \rightarrow N$.
Such degeneracies can be induced by the gate voltage $U_g$ and appear
in the simplest possible model (constant interaction model) periodic in
$U_g$. In SWNTs Coulomb blockade has been observed by several groups
before.\cite{CB_SWNT}
Until now, it was believed that MWNTs do not display
Coulomb blockade because of low contact resistances.
Recently, we have observed clear signatures
of single-electron charging in a \mbox{$300$\,nm}
long MWNT (distance between the contacts) with
reasonably low-ohmic contacts
(i.e. contact resistance $\ltequiv h/2e^2$).
An example is shown
in Fig.~4 \mbox{($T=270$\,mK)}.
In (a), measured for $B=0$,
a regular sequence  is seen
consisting of a large `diamond' followed by three smaller
ones.\footnote{\,For other gate voltages the pattern can look quite
different, not resembling SET effects at all.}
This pattern suggests a four-fold degeneracy of the
single-electron states which we assign to spin degeneracy
and an additional twofold orbital degeneracy.
The latter may be caused by  a peculiar degeneracy inherent
to the graphite lattice, the so-called
K, K$^{\prime}$  degeneracy which originates from
the presence of two identical carbon atoms per unit cell.
If this interpretation is correct, this is quite amusing,
because this basic feature has up to now not been
seen in SWNTs.

Fig.~4b shows $dI/dV$ measured at \mbox{$B=4$\,T}. Now,
an irregular sequence is observed. It is well known that the
K, K$^{\prime}$  degeneracy is lifted in magnetic-field.

On the one hand, we do see clear signatures of Coulomb
blockade suggesting low transparent contacts. On the
other hand, the measured two-terminal resistance is
low, of order of
$h/2e^2$, suggesting highly (or intermediate) transparent
contacts. Therefore, higher-order tunneling processes can
play an important role in the electrical transport.
A well known higher-order tunneling process
is the Kondo effect, i.e. the dynamic screening
of the spin on the dot due to exchange-correlation effects
between the dot state and the electrodes.
We observe the Kondo effect in a MWNT for the
first time, see the features in Fig.~4., which are highlighted
by arrows. At $V=0$ and within the region of
Coulomb blockade there are two `ridges' of enhanced
conductance. For the simplest filling scheme, i.e.
$S=0\rightarrow 1/2\rightarrow 0 \dots$, these ridges are
expected to form in every second  `diamond' where
the spin is $S=1/2$. The Kondo effect has
been observed before in SWNTs.\cite{Nygard}

The data in Fig.~4 allow to extract the single-particle
level spacing $\delta E$. The large diamonds are large because
an electron has to be added to a new level. The addition energy
is therefore the sum of charging energy $E_c$ and $\delta E$.
Due to the fourfold degeneracy, the addition energy for the
next three electrons is only $E_c$. Comparing the energies
leads to a mean level spacing of \mbox{$\delta E \approx 0.4$\,meV}.
The $0$d level spacing
is predicted to be \mbox{$\delta E = h v_F/2L$} for one mode,
yielding \mbox{$5.5$\,meV} (\mbox{$L=300$\,nm}). The number of modes
participating is therefore \mbox{$M=5.5/0.4\approx 14$}. Again we see
that there are more than $2$ modes
occupied in MWNTs.\footnote{\,Note, that we have taken \mbox{$v_F=8\cdot 10^5$\,m/s}
to be constant in this estimate. This might be wrong because of the dispersion of higher
subbands, lowering $v_F$.}

Since the measured two-terminal conductance
$G\sim 2e^2/h$,\footnote{\,$G\approx 2e^2/h$ is a
typical value for all our MWNT devices with contact separation
$L\approx 300$\,nm.} the
mean transmission probability $T$ per mode is only $0.1$ (here, we have
taken $M=10$). If all modes would have $T=0.1$, co-tunneling effects
like the Kondo effects could safely be neglected. Hence, there
must be a distribution of transmission eigenvalues similar to
that in many-mode diffusive wires. For the latter, random-matrix shows
that the distribution is bimodal. Although we are dealing with a
very small ensemble (only 10 modes), the distribution of transmission
values is most likely bimodal, too. Though there are $\approx 10$ modes
available to accommodate charge, charge transport to the contacts
is determined by only one mode (or a very small number of modes).
If the mode with highest transmission probability has the appropriate
$T$ (not too low, but also not too high), the Kondo effect will
show up.\cite{Nygard2}
If $T$ is larger,
the observed conductance pattern can no longer be assigned
to co-tunneling SET effects but rather reflects Fabry-Perot type
interference.\cite{FabryPerot}

\section{Conclusion}

Multi-wall nanotubes in air are hole doped, most likely induced
by the adsorption of water. Consequently, there are more than
two $1$d-modes
occupied. A typical number is $10$, not counting
spin degeneracy. Estimates for the scattering
mean-free path $l$ vary widely in the literature and also among
our own experiments. However, surprisingly many experiments
suggests that the $l$ is of order of the circumference
of the tube. This may be a coincidence, but might also
have a deeper reason: It was pointed out very early
by White {\it et al.}\cite{White}
that the mean-free path should grow with tube diameter $d$
according to $l \propto d$, provided the number of scatterers
per unit length is constant.
On the other hand, if we assume that the source of scattering is
a direct consequence of doping, the scattering density is
proportional to the doping density which for a given
environment can be taken to be constant. Therefore, the number
of dopants per unit length (or the number of scatterers) grows linearly with
$d$. The combination of both effects results in a diameter
independent mean-free path. Another argument for a diameter-independent
mean-free path in MWNTs was put forward by Egger and Gogolin.\cite{Egger}

At low temperature, single-electron tunneling (SET) effects have
been observed in MWNTs with low-ohmic contacts, i.e. \mbox{$\ltequiv 10$\,k$\Omega$}.
This has allowed to
determine the single electron level spacing which appears to be consistent
with the estimated number of occupied modes deduced from electrochemical
gating. The presence of co-tunneling points to high transmission
at the contacts. Even stronger: the prevailing absence of SET effects
in MWNTs with $G\sim 2e^2/h$ suggests that at least one highly transmissive channel
is present at low temperature.

\end{document}